\documentclass[3p]{elsarticle}

\usepackage{setspace}
\usepackage{hyperref}
\usepackage{bm}
\usepackage{natbib}
\usepackage{amsmath}
\usepackage{graphicx}
\usepackage{subcaption}
\usepackage{footnpag}			      	
\usepackage{float}
\usepackage{multirow}
\graphicspath{ {Figures/} }

\setcitestyle{authoryear,round}

\newcommand{\bfm}[1]{\mbox{\boldmath{$#1$}}}

\begin{document}

\begin{frontmatter}

\title{Spin State Evolution of (99942) Apophis during its 2029 Earth Encounter}

\author{Conor J. Benson}

\author{Daniel J. Scheeres}
\address{University of Colorado Boulder, 3775 Discovery Drive, Boulder, CO 80303, USA}

\author{Marina Brozovi\'{c}}

\author{Steven Chesley}
\address{Jet Propulsion Laboratory, California Institute of Technology, 4800 Oak Grove Drive, Pasadena, CA 91109, USA}

\author{Petr Pravec}

\author{Petr Scheirich}
\address{Astronomical Institute, Academy of Sciences of the Czech Republic, Fri\v{c}ova 1, CZ-25165 Ond\v{r}ejov, Czech Republic}

\begin{abstract}

We explore the effects of the 2029 Earth encounter on asteroid (99942) Apophis' non-principal axis spin state, leveraging refined orbit, spin state, and inertia information provided by more recent optical and radar observations. Propagating the asteroids' coupled orbit and rigid body attitude dynamics through the flyby, we present the range of possible post-flyby spin states. These spin state distributions will be valuable for planning Apophis observation campaigns and spacecraft missions, most notably OSIRIS-APEX. The simulations indicate that gravitationally induced changes to the asteroid's tumbling periods and rotational angular momentum direction (pole) will likely be significant and measurable. For the current spin state and inertia estimates and their uncertainties, Apophis is likely to remain in a short axis mode (SAM) tumbling state but its effective spin rate could halve or double. Its pole is likely to shift by 10 degrees or more and increase in longitude while moving closer to the ecliptic plane. These spin state changes are very sensitive to the asteroid's close approach attitude and mass distribution. With ground-based tracking of the asteroid's spin state through the encounter, this sensitivity will help refine mass distribution knowledge.  We also discuss the implications of this abrupt spin state alteration for Apophis' Yarkovsky acceleration and geophysical properties, identifying possible pathways for surface and internal changes, most notably if Apophis is a contact binary. Comparison of the pre and post-flyby inertia estimates obtained from the ground-based observations will help assess the extent of possible geophysical changes. 

\end{abstract}

\begin{keyword}
Asteroids, dynamics; Asteroids, rotation; Near-Earth objects
\end{keyword}

\end{frontmatter}

\section{Introduction}

On April 13, 2029 asteroid (99942) Apophis will make a close approach to the Earth, coming within 6 Earth radii from the geocenter. This unprecedented flyby offers a unique opportunity to learn about small body evolution and has been highlighted in the Planetary Science Decadal Survey \citep{decadal2023}. It is an event that will drive much planning and analysis. Of particular interest is NASA's OSIRIS-APEX mission, which is set to rendezvous with Apophis several months after the asteroid's 2029 Earth encounter \citep{orex_apophis}. The close approach will have a significant impact on two aspects of Apophis’ dynamic state, its orbit and its rotation. In this paper we revisit an analysis by \cite{scheeresapophis} conducted shortly after Apophis' discovery in 2004 that studied the range of possible spin states the asteroid could have following its close approach flyby. Given the limited observations up to that point, this study was subject to large uncertainties in the close approach distance and asteroid physical properties. From preliminary light curve analysis, the asteroid was modeled as a triaxial ellipsoid in uniform rotation about its maximum inertia axis. In our current analysis, we take advantage of Apophis' greatly constrained orbit, spin state, and shape made possible by extensive subsequent observations. Specifically, we draw from much richer knowledge of its non-principal axis rotation state and shape obtained from optical \citep{pravecapophis} and radar \citep{brozovicapophis} measurements. These important refinements greatly improve the modeling of the effects of Apophis' Earth encounter and will provide a realistic range of post-flyby spin states that may be expected.

This analysis supports several important scientific aspects. First, the flyby will provide insight into the asteroid's mass distribution and interior based on the observed changes in its spin state. When the actual moments of inertia are compared to those derived from the constant density asteroid shape model, the extent of Apophis' density homogeneity can be explored. Discrepancies between the shape-derived and actual moments of inertia could indicate density variations across the asteroid's surface and/or interior. The convex shape model solution of \cite{pravecapophis} shows some discrepancy between these inertias. But it is very possible that these differences are due to surface concavities that cannot be resolved with current optical observations. The radar observations suggest that Apophis has a bi-lobed shape \citep{brozovicapophis}, supporting this hypothesis. Improvements in shape and spin state provided by future observations, particularly those around the 2029 encounter, should allow for further refinements to the shape, flyby-induced evolution, and mass distribution. Detailed knowledge of Apophis' spin state and mass distribution will be crucial for planning and operating in-situ missions such as OSIRIS-APEX. 

These spin state simulation results will allow also us to predict the range of surface accelerations and global stresses that will be placed across the body during its closest approach. Previous work by \cite{scheeresapophis}, \cite{demartini2019using}, and  \cite{hirabayashiapophis} explored the geophysical implications of the flyby. All three studies make simplifying assumptions about the pre-encounter spin state and its uncertainty. Both \cite{scheeresapophis} and \cite{demartini2019using} do not account for the asteroid's non-principal axis rotation. \cite{hirabayashiapophis} consider the full \cite{pravecapophis} spin state solution but do not account for current uncertainties in Apophis' tumbling periods and rotational angular momentum direction (pole) and the resulting increase in possible close encounter attitudes. \cite{souchay2018} explore the effects of the encounter on Apophis' pole direction using the \cite{pravecapophis} solutions and their uncertainties, but they assume the asteroid is uniformly rotating about its maximum inertia axis. Furthermore, \cite{souchay2018} do not explore how the asteroid's tumbling periods will change due to the gravitational torques. In this work, we conduct rigid body dynamical modeling that accounts for both the asteroid's non-principal axis rotation and initial condition uncertainties. We present post-flyby distributions for both the pole and tumbling periods, providing a complete picture of possible encounter outcomes from which to address geophysical implications. These predictions may enable more precise designs of any measurements that may be performed by visiting spacecraft. Finally, the flyby-induced spin state change may alter the asteroid's Yarkovsky acceleration and subsequent Earth encounter predictions. 

In this paper, we outline the current Apophis spin state estimates, propagate these states through the flyby, present the resulting post-flyby spin state distributions, and finally discuss implications for Apophis' flyby-induced geophysical evolution and the Yarkovsky effect.

\section{Current Spin State Estimates}

\begin{figure}[H]
    \centering
    \includegraphics[width=4in]{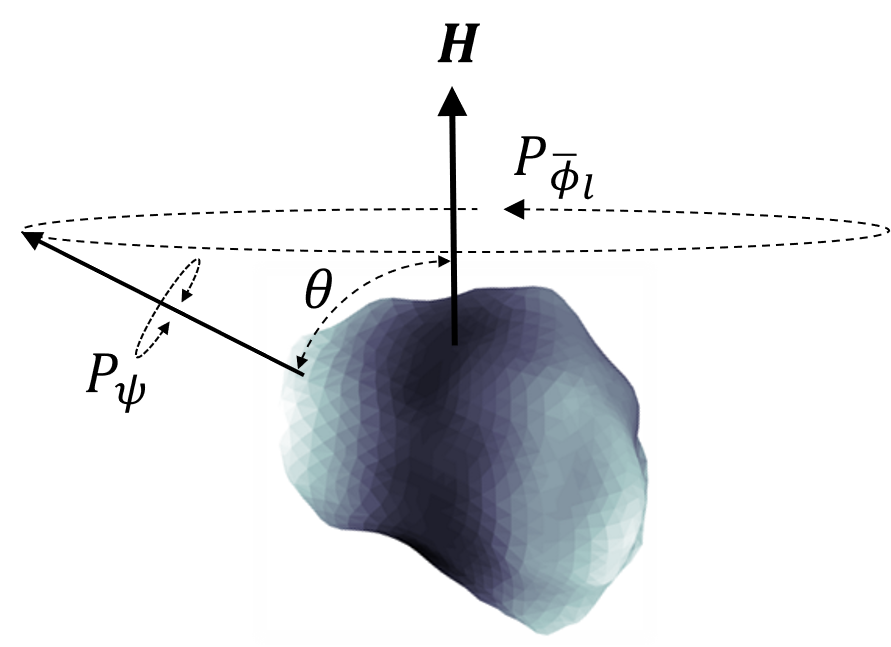}
    \caption{Long axis tumbling period convention for Apophis}
    \label{fig:apophis_periods}
\end{figure}

The Apophis spin state and shape model solutions provided by \cite{pravecapophis} and \cite{brozovicapophis} are both considered in our analysis. From here on, these will be referred to as the ``photometric" and ``radar" solutions respectively. Table~\ref{tab:preflybysol} lists these solutions. Here, $P_{\bar{\phi}_s}$ is the average precession period of the asteroid's maximum inertia (short) axis around the rotational angular momentum vector $\bm{H}$ \citep{sm2015}. This is the short axis period convention assumed in \cite{pravecapophis}. Alternatively, we can use the long axis convention where $P_{\bar{\phi}_l}$ is the average precession period of the minumum inertia (long) axis about $\bm{H}$ \citep{sa1991}. Given Apophis' elongated shape, it can be more intuitive to track motion of the long axis. $P_{\bar{\phi}_l}$ also tends to have the largest associated amplitude in Apophis light curves \citep{pravecapophis}. The other tumbling period, $P_\psi$, is the rotation period about the short or long axis in the corresponding convention. Since $P_\psi$ is equal to the circulation period of the asteroid's angular velocity vector $\bm{\omega}$ in the body-fixed frame, its value is the same for both short and long axis conventions \citep{sm2015}. A schematic of the tumbling periods for the long axis convention is shown in Figure~\ref{fig:apophis_periods} where $\theta$ is the time-varying nutation angle between the long axis and $\bm{H}$. 

The three tumbling periods are related by the following equation \citep{sm2015},

\begin{equation}
\frac{1}{P_{\bar{\phi}_s}} = \frac{1}{P_{\bar{\phi}_l}} + \frac{1}{P_{\psi}}
\label{eq:pphis}
\end{equation}

In Table~\ref{tab:preflybysol}, $\lambda$ and $\beta$ are the J2000 ecliptic longitude and latitude of $\bm{H}$. Finally, denoting the three principal moments of inertia as $I_l\;{\leq}\;I_i\;{\leq}\;I_s$, we provide the two inertia ratios $I_l$/$I_s$ and $I_i$/$I_s$. Uncertainties in the photometric solution quantities are given to the 3$\sigma$ confidence level. For the photometric solution, the nominal $\lambda$ and $\beta$ values are given while \cite{pravecapophis} provide the range of admissible rotational angular momentum directions in Figure 4 of their paper. Throughout this work we refer to the angular momentum direction as the ``pole". Due to the limited resolution of the 2012 - 2013 Apophis radar data given the relatively large 0.1 AU encounter distance, \cite{brozovicapophis} were unable to improve on the photometric solution uncertainties. So the radar solution is subject to similar uncertainties. See \cite{brozovicapophis} and its supplemental material for more insight on the radar solution uncertainties. 

\begin{table}[H]
\centering
\caption{Apophis nominal pre-flyby spin state solutions. Both correspond to short axis mode (SAM) spin states.}
\begin{tabular}{| c | c | c | c | c | c | c | c |}
\hline
Model & $P_{\bar{\phi}_s}$ (hr) & $P_{\bar{\phi}_l}$ (hr) & $P_{\psi}$ (hr) & $\lambda$ & $\beta$ & $I_l$/$I_s$ & $I_i$/$I_s$\\
 \hline
Photometric & 27.38$\;\pm\;{0.07}$ & 30.56 $\;\pm\;{0.01}$ & 263$\;\pm\;{6}$ & 250$^{\circ}$ & -75$^{\circ}$ & 0.61$^{+.11}_{-.08}$ & 0.965$^{+.009}_{-.015}$\\
\hline
Radar & 27.45 & 30.62 & 265.7 & 247$^{\circ}$ & -59$^{\circ}$ & 0.73 & 0.95\\
\hline
\end{tabular}
\label{tab:preflybysol}
\end{table}

\section{Flyby Simulations}

\subsection{Approach}
We model Apophis as a rigid body and use a second degree and order gravity field derived from the Table~\ref{tab:preflybysol} moment of inertia ratios \citep{scheeresbook}. Assuming a point mass Earth given the rapid fall off of higher order perturbations with distance, we numerically integrate the asteroid's coupled orbit and attitude dynamics over four days centered on the 2029 encounter. We conducted very similar flyby analysis for the 2013 Earth encounter of the $\sim$40 m asteroid (367943) Duende \citep{bensonda14}. Duende approached within $\sim$34,000 km of Earth's center, about 4,000 km closer than Apophis' predicted distance in 2029. Like Apophis, Duende is tumbling and elongated, with an approximately 2:1 long-short axis ratio. This elongation allows for generation of significant gravitational torques and flyby simulations indicated that Duende's spin likely changed significantly. Supporting this, the best-fit post-flyby light curve tumbling periods were found to be inconsistent with the pre-flyby light curve \citep{moskovitzda14, bensonda14}. For both Duende and Apophis, simulations show that almost all spin state changes occur within hours of closest approach. For example, at five hours before or after closest approach, the torque is only 3\% of the closest approach value. This is due to the rapid $1/R^3$ fall-off in gravitational torque with distance. Returning to details of the current Apophis analysis, we use JPL Horizons ephemerides for the orbit initial conditions. A March 2019 orbit solution with a nominal close approach radius of 37730 km is used. We note that this is $\sim$280 km lower than the most recent nominal solution of 38012 km. This yields a maximum difference in gravitational torques of only $\sim$2\% which quickly decays to zero away from closest approach. So differences between the resulting post-flyby state distributions for these orbital solutions will be negligible.  Given current uncertainty in Apophis' two tumbling periods, the 2029 close approach attitude ($\phi_o$) and angular velocity ($\tau_o$) phasing around $\bm{H}$ \citep{scheeresbook, bensonda14} are unknown. There is also uncertainty in the asteroid's current pole direction and moments of inertia. Given this overall uncertainty, we conduct an ensemble of 50,000 runs to map out the range of possible flyby outcomes. We assume the nominal photometric periods given their small relative uncertainty and the limited influence the encounter periods were found to have on flyby outcomes. To account for the phasing uncertainty associated with the periods, we uniformly sample $\phi_o$ and $\tau_o$ over all possible values.  We also uniformly sample the pole direction over the \cite{pravecapophis} admissible region (which contains the nominal radar-derived pole). For the inertia ratios, we uniformly sample $I_l/I_s\;{\in}\;[0.53,0.73]$ and $I_i/I_s\;{\in}\;[0.95,0.974]$ for consistency with both the photometric and radar solutions. Inertia pairs where both $I_l/I_s\;{\rightarrow}\;0.73$ and $I_i/I_s\;{\rightarrow}\;0.974$ are not dynamically viable in torque-free rigid body rotation for the nominal tumbling periods \citep{sa1991} and are therefore excluded. 

\subsection{Results}

\begin{figure}[H]
	\centering
    \includegraphics[width=5.5in]{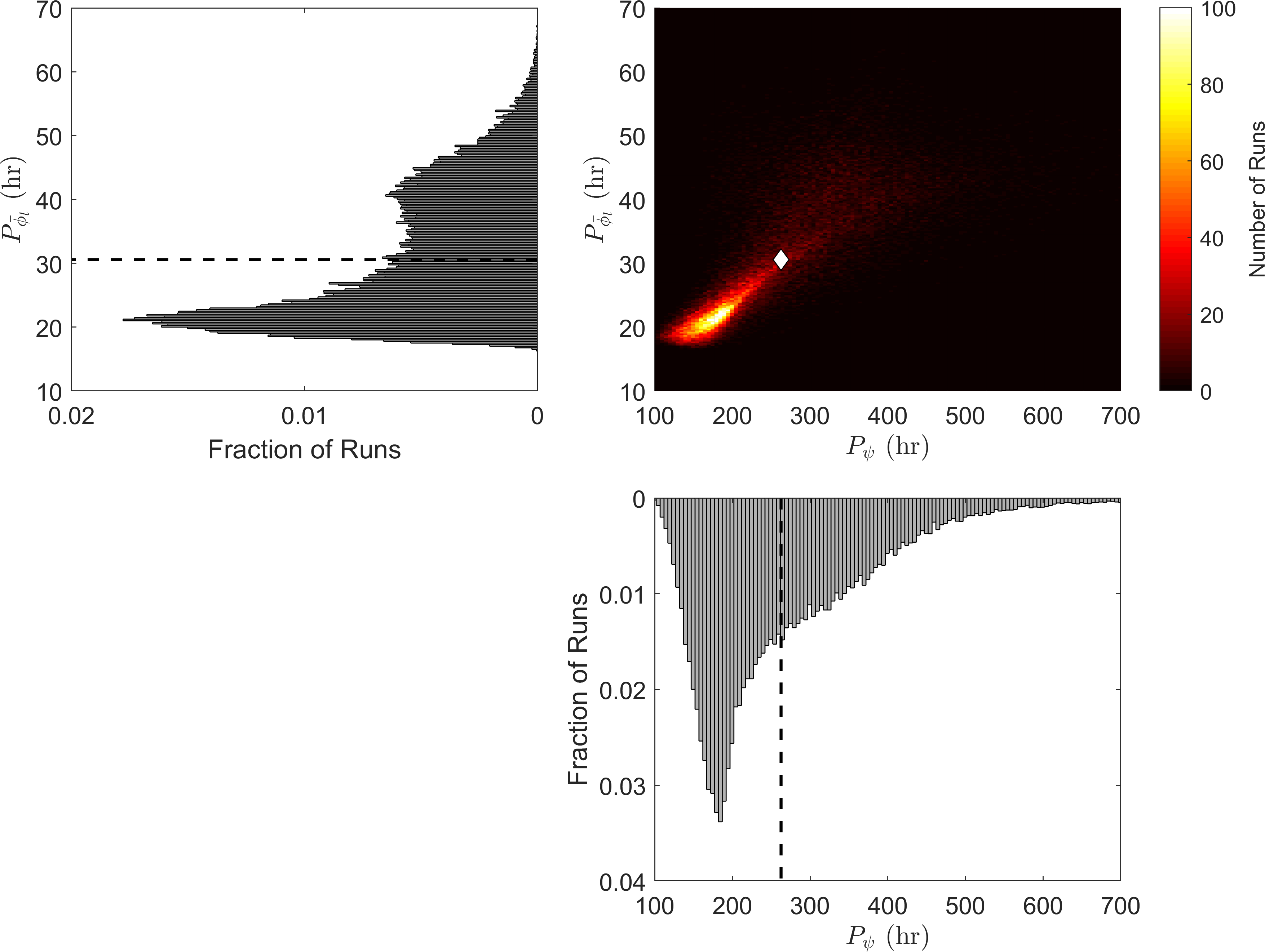}
    \caption{Apophis post-flyby tumbling period distributions (50,000 runs). The dashed lines and white diamond denote the pre-flyby periods.}
    \label{fig:pposthist}
\end{figure}

We first discuss the post-flyby $P_{\bar{\phi}_l}$ and $P_{\psi}$ distributions. These are provided in Figure~\ref{fig:pposthist}. Histograms are shown for each of the tumbling periods separately and also for the combined distribution. The dashed lines and white diamond represent the pre-flyby values. There is wide dispersion in the distributions for both periods. $P_{\bar{\phi}_l}$ ranges from 16.4 - 67.2 hr with a mean of 30.9 hr (30.56 hr pre-flyby). The $P_{\bar{\phi}_l}$ distribution is also bimodal with a highest probability value of $\sim$21 hr. This bimodality can be explained by the gravitational torque structure. $P_{\bar{\phi}_l}$ will remain relatively constant if the asteroid long axis is nearly aligned with or perpendicular to the Earth direction at closest approach (the orientations with the smallest gravitational torque). Since we uniformly sample over $\phi_o$ from 0$^{\circ}$ - 360$^{\circ}$, relatively few runs yield these torque-minimizing geometries. For most of the sampled orientations, the long axis is offset from the parallel/perpendicular directions at closest approach, resulting in a net acceleration that causes either an increase or decrease in $P_{\bar{\phi}_l}$. For $P_{\psi}$, post-flyby values range from 95.8 - 1686.9 hr (70.3 days) with a mean of 266.0 hr (263 hr pre-flyby) and a highest probability value of $\sim$185 hr. There is strong positive correlation between the two periods. In other words, runs with smaller $P_{\bar{\phi}_l}$ tend to have smaller $P_{\psi}$ and vice versa. In 55.8\% of runs, $P_{\bar{\phi}_l}$ decreases through the flyby corresponding to faster precession of the long axis about $\bm{H}$.  In 59.3\% of runs, $P_{\psi}$ decreases through the flyby. Of particular note is the concentration of outcomes in the lower left of the combined distribution plot. This hotspot, defined by decreases in both periods, contains 51.2\% of runs. The hotspot persisted when only the close approach phasing was varied, assuming nominal pole and inertias nominal values. This indicates the hotspot is dictated primarily by the flyby geometry and can be expected to persist for any viable pole or inertia ratios. While the hotspot is prominent, it should be noted that the post-flyby periods are almost as likely to lie outside the hotspot given the current uncertainties. 

We can also consider the post-flyby states in terms of their effective spin rate $\omega_e=2T/H$ and dynamic moment of inertia $I_d=H^2/2T$ where $H$ and $T$ are the rotational angular momentum magnitude and kinetic energy \citep{scheeresbook,bensonda14}.  The effective spin rate $\omega_e$ is proportional to the angular velocity vector $\bm{\omega}$ and inversely proportional to the two tumbling periods. Therefore, $\omega_e$ is a convenient indicator for the asteroid's overall spin rate. $I_d$ determines how far the asteroid is into tumbling.  $I_d=I_l$ and $I_d=I_s$ correspond to uniform rotation about the long and short axes respectively. For $I_i<I_d<I_s$, the asteroid is in short axis mode (SAM) rotation where $\bm{\omega}$ precesses about the short axis. For SAMs, the asteroid rocks back and forth about its long axis while this axis precesses about $\bm{H}$.  For $I_l<I_d<I_i$, the asteroid is in a long axis mode (LAM) where $\bm{\omega}$ instead precesses about the long axis. For LAMs, the asteroid continuously rotates about its long axis while precessing about $\bm{H}$. Finally, for $I_d=I_i$, the asteroid is in uniform rotation about the intermediate axis or evolving along the separatrix between LAM and SAM states. To properly compare the tumbling level for the ensemble of runs with different inertias, we use the scaled dynamic moment of inertia $\tilde{I_d}$. For SAMs, $\tilde{I_d}=(I_d-I_i)/(I_s-I_i)$. For LAMs, $\tilde{I_d}=(I_d-I_i)/(I_i-I_l)$. So $-1\leq\tilde{I_d}\leq1$ with the extremal values indicating uniform long/short axis rotation respectively and 0 indicating intermediate axis rotation or motion along the separatrix.

\begin{figure}[H]
    \centering
    \includegraphics[height=3in]{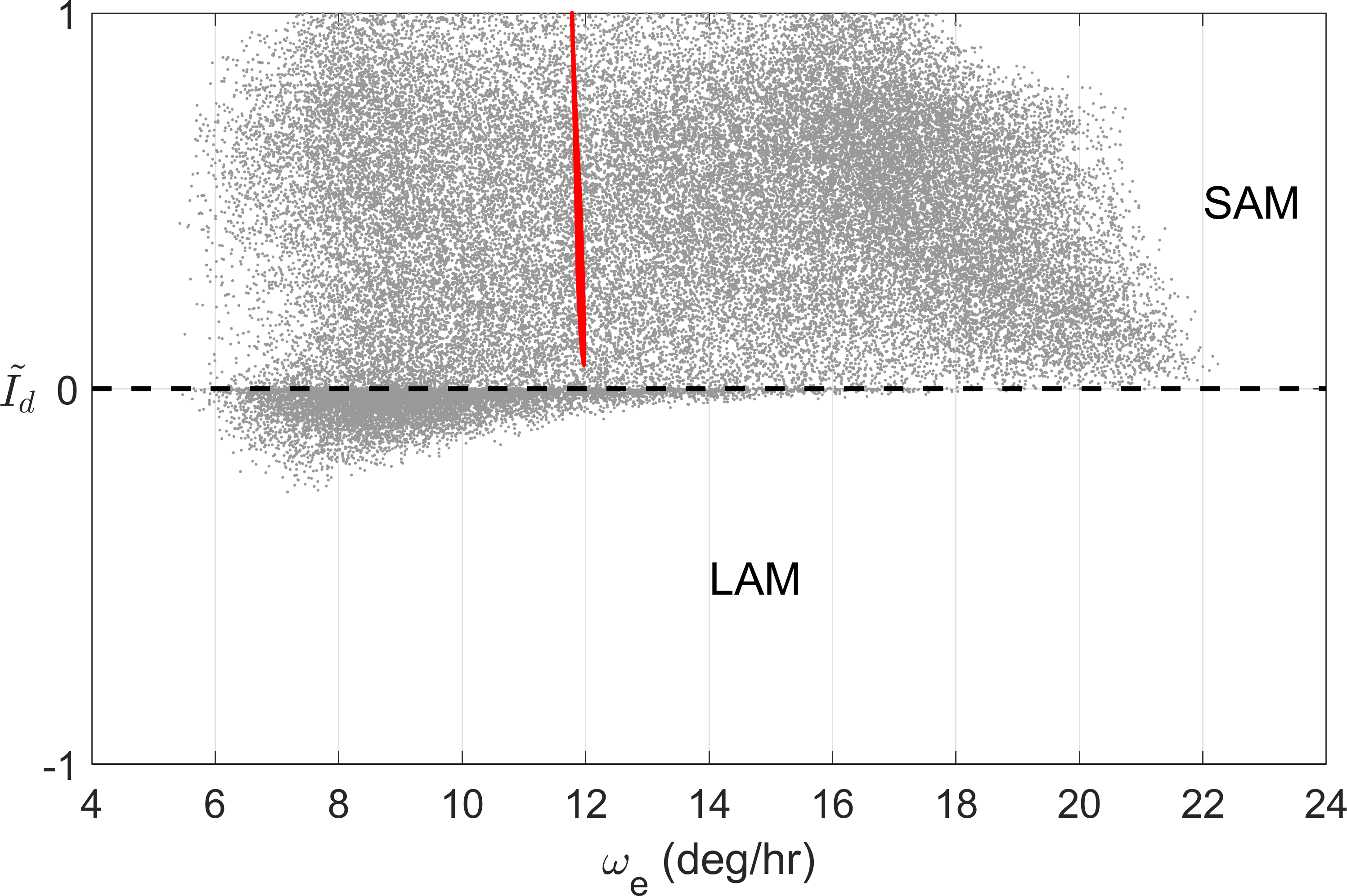}
    \caption{Corresponding post-flyby distributions of Apophis' effective spin rate $\omega_e$ and normalized dynamic moment of inertia $\tilde{I_d}$. The dashed line denotes the separatrix between LAM and SAM states. The red region denotes the uniformly sampled pre-flyby initial conditions.}
    \label{fig:idwlpost}
\end{figure}

Figure~\ref{fig:idwlpost} shows the post-flyby ($\tilde{I_d}$, $\omega_e$) distribution. The region in red denotes the pre-flyby values. Over the range of possible outcomes, $\omega_e$ roughly halves or doubles compared to the pre-flyby value. In 55.8\% of cases, $\omega_e$ increases pre to post-flyby. $\tilde{I_d}$ on the other hand does not greatly exceed its pre-flyby range with only 17.3\% of runs transitioning to LAM. With constant $I_d$ corresponding to constant $P_{\psi}/P_{\bar{\phi}_{l,s}}$ \citep{scheeresbook,bensonda14}, variation mostly in $\omega_e$ is consistent with the positive period correlation (i.e. roughly constant distribution slope) in Figure~\ref{fig:pposthist}.

\begin{figure}[H]
    \centering
    \includegraphics[height=3in]{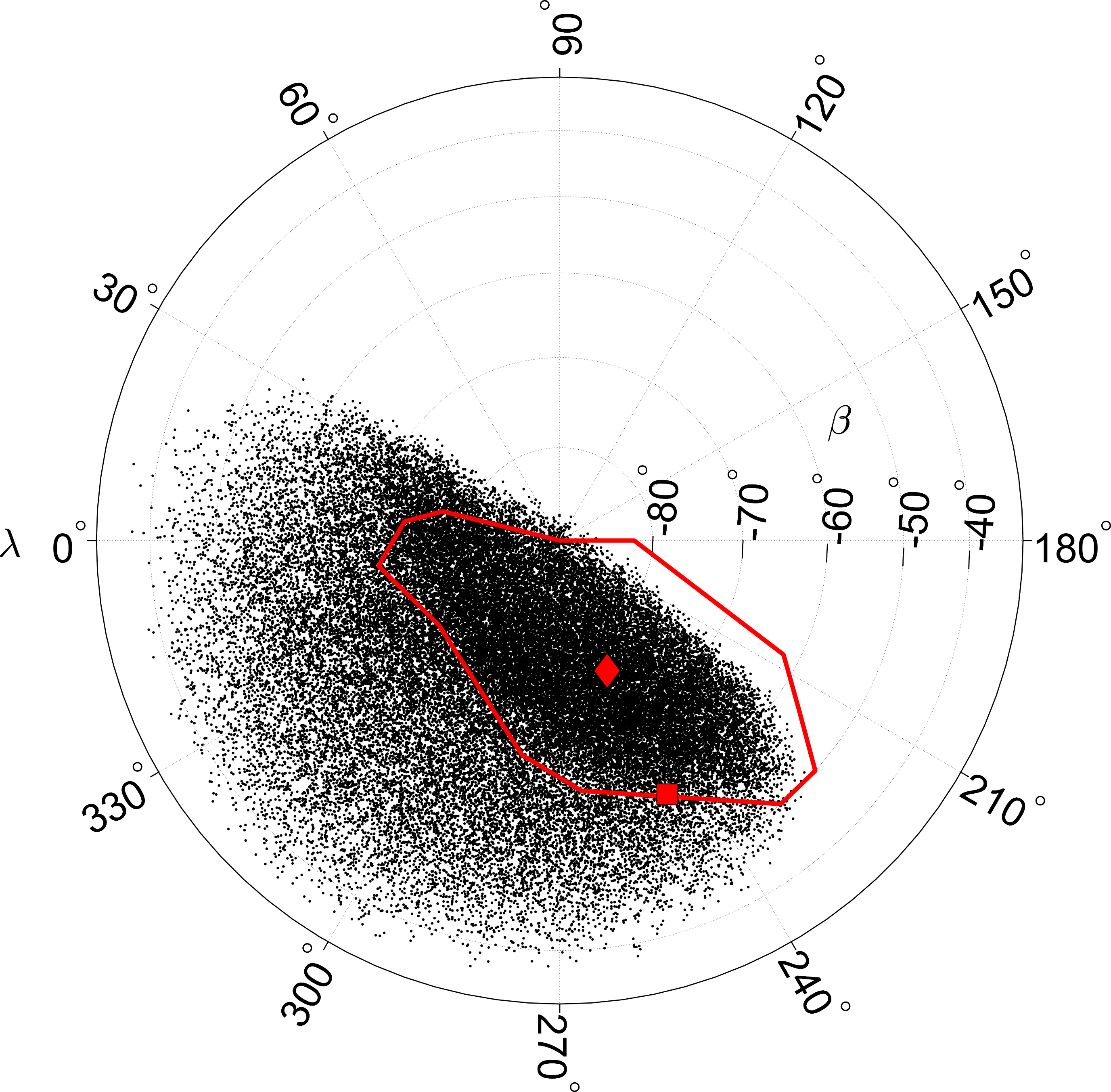}
    \caption{Apophis post-flyby rotational angular momentum directions in terms of J2000 ecliptic longitude $\lambda$ and latitude $\beta$. The red diamond and square denote the nominal photometric and radar pre-flyby directions respectively. The region bounded in red denotes the uniformly sampled admissible pre-flyby pole region from \cite{pravecapophis}.}
    \label{fig:postpole}
\end{figure}

\begin{figure}[H]
    \centering
    \includegraphics[height=3in]{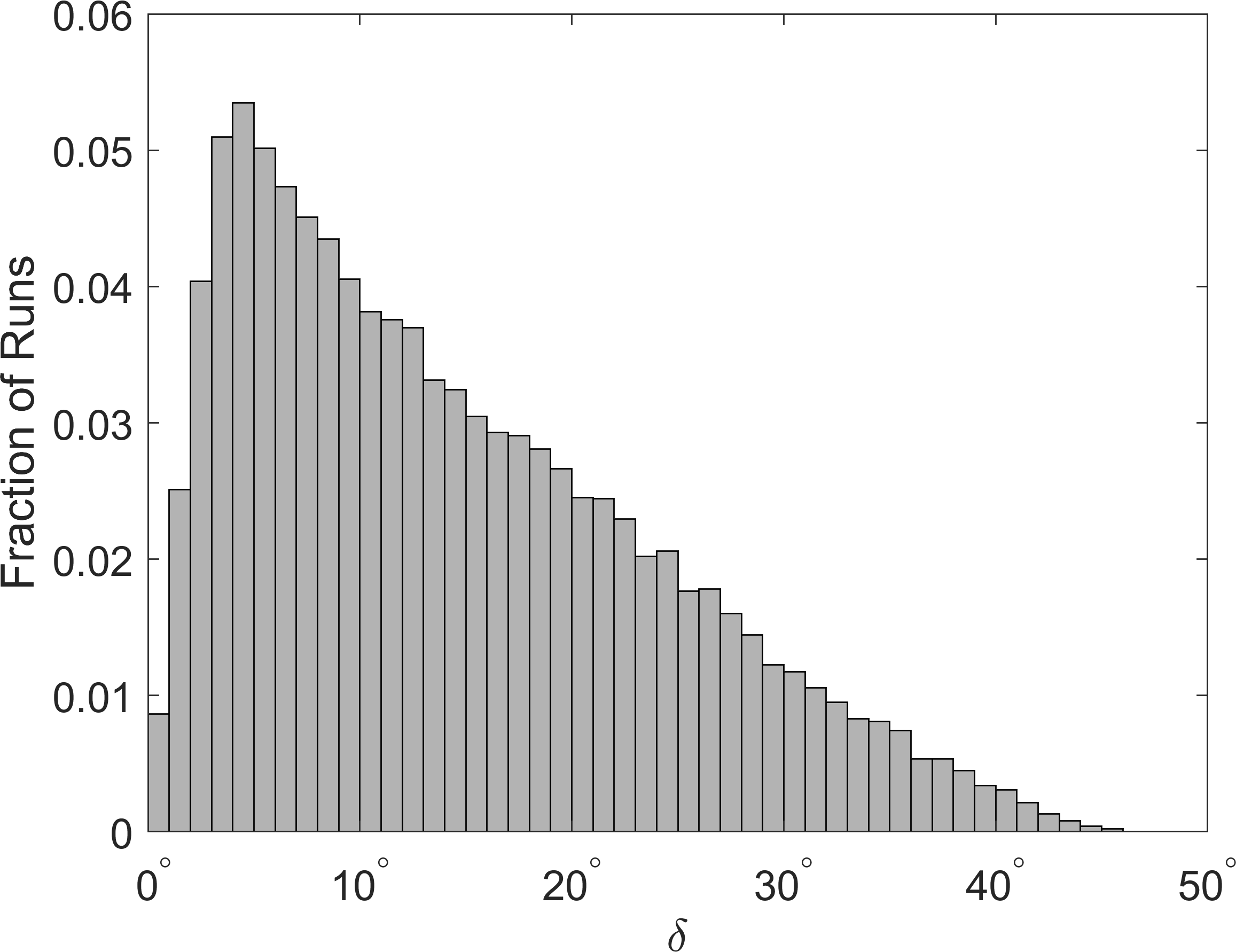}
    \caption{Histogram of the angle $\delta$ between the pre and post-flyby angular momentum directions.}
    \label{fig:postpole_deltahist}
\end{figure}

Figure~\ref{fig:postpole} shows the post-flyby pole distribution in the J2000 ecliptic frame. The red boundary is the admissible pre-flyby pole region from \cite{pravecapophis}. Again, the pre-flyby poles were uniformly sampled from this bounded region. The red diamond and square denote the nominal photometric and radar pre-flyby directions respectively. There is a notable leftward trend in pole motion with 93.5\% of runs having an increased post-flyby pole longitude. This structure is due to Apophis' particular flyby geometry and the fact that a significant component of the gravitational torque acts in a direction that would move the long axis towards the instantaneous earth line, rather than away from it. The change in rotational angular momentum is along this torque direction, so solutions are grouped to one side of the southern hemisphere. Considering pole latitude, 55.3\% of poles move closer to the ecliptic plane. The complementary Figure~\ref{fig:postpole_deltahist} shows the distribution of the angular separation $\delta$ between the pre and post-flyby poles. Here, the mean value is 14.5$^{\circ}$ with $\sim$5$^{\circ}$ being the highest probability outcome. Values as high as 45.8$^{\circ}$ are observed. 

\begin{figure}[H]
    \centering
    \subcaptionbox{histogram of closest approach angular acceleration magnitudes}{\vspace{0.25in}
\includegraphics[height=2.5in]{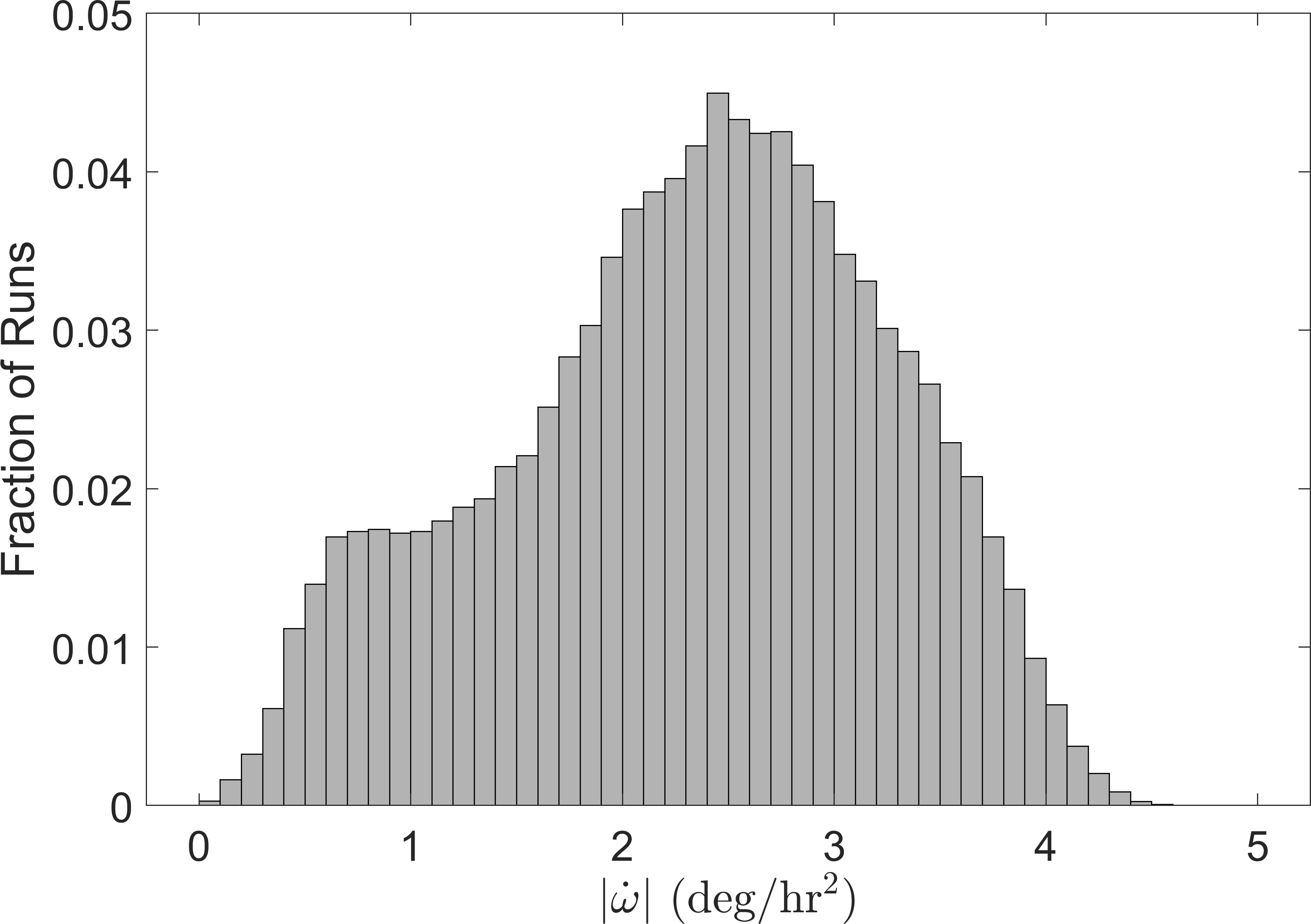}}\hfill
    \subcaptionbox{angular acceleration vectors along the short ($s$), intermediate ($i$), and long ($l$) axes with the grey regions denoting projections onto the principal planes}{\includegraphics[height=3in]{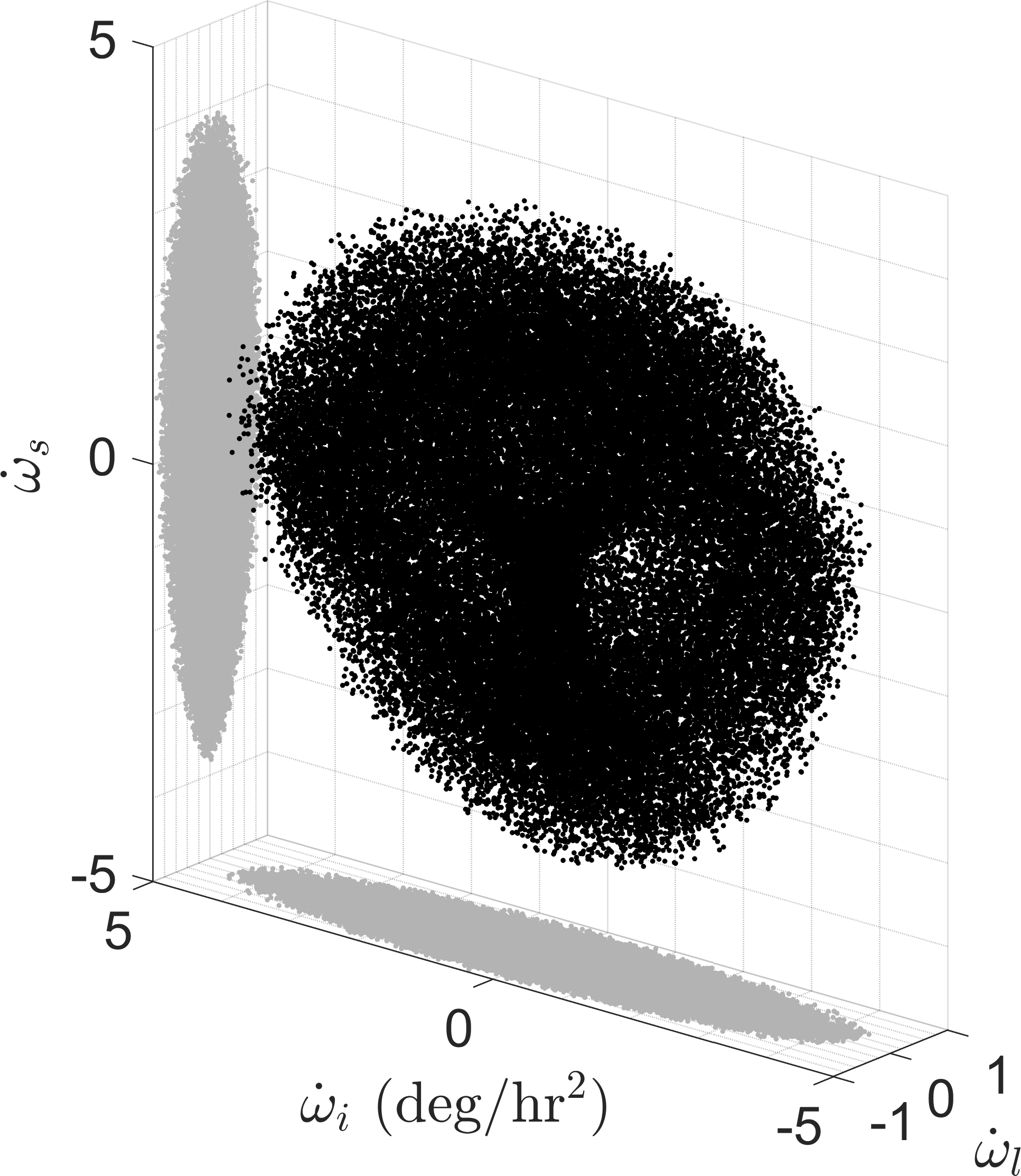}}
    \caption{Apophis closest approach angular acceleration distributions}
    \label{fig:wdotca}
\end{figure}

With implications for possible geophysical changes, Figure~\ref{fig:wdotca}a shows the histogram of the angular acceleration magnitude at closest approach, where accelerations tend to be largest. The maximum acceleration magnitude is 4.5 deg/hr$^2$ (6$\times$10$^{-9}$ rad/s$^2$) and the highest probability outcome is 2.4 deg/hr$^2$ (3.2$\times$10$^{-9}$ rad/s$^2$). For reference, in torque-free rotation, the maximum angular acceleration for the nominal photometric spin state is $\sim$0.4 deg/hr$^2$ (5.4$\times$10$^{-10}$ rad/s$^2$), roughly one order of magnitude smaller. Figure~\ref{fig:wdotca}b shows the corresponding closest approach acceleration vectors plotted along the asteroid's long, intermediate, and short axes for all 50,000 runs. These accelerations map out a relatively thin ``disk" with maximum accelerations of $\pm$1 deg/hr$^2$ about the long axis and values roughly 5 times that for the intermediate and short axes. Accelerations are largest about the intermediate ($i$) and short ($s$) axes given the longer lever arms perpendicular to these axes. Maximum accelerations are slightly larger about the intermediate axis, likely given its smaller inertia. 

\begin{figure}[H]
    \centering
    \subcaptionbox{varying $\overline{I_l}=I_l/I_s$}{\includegraphics[width=3.2in]{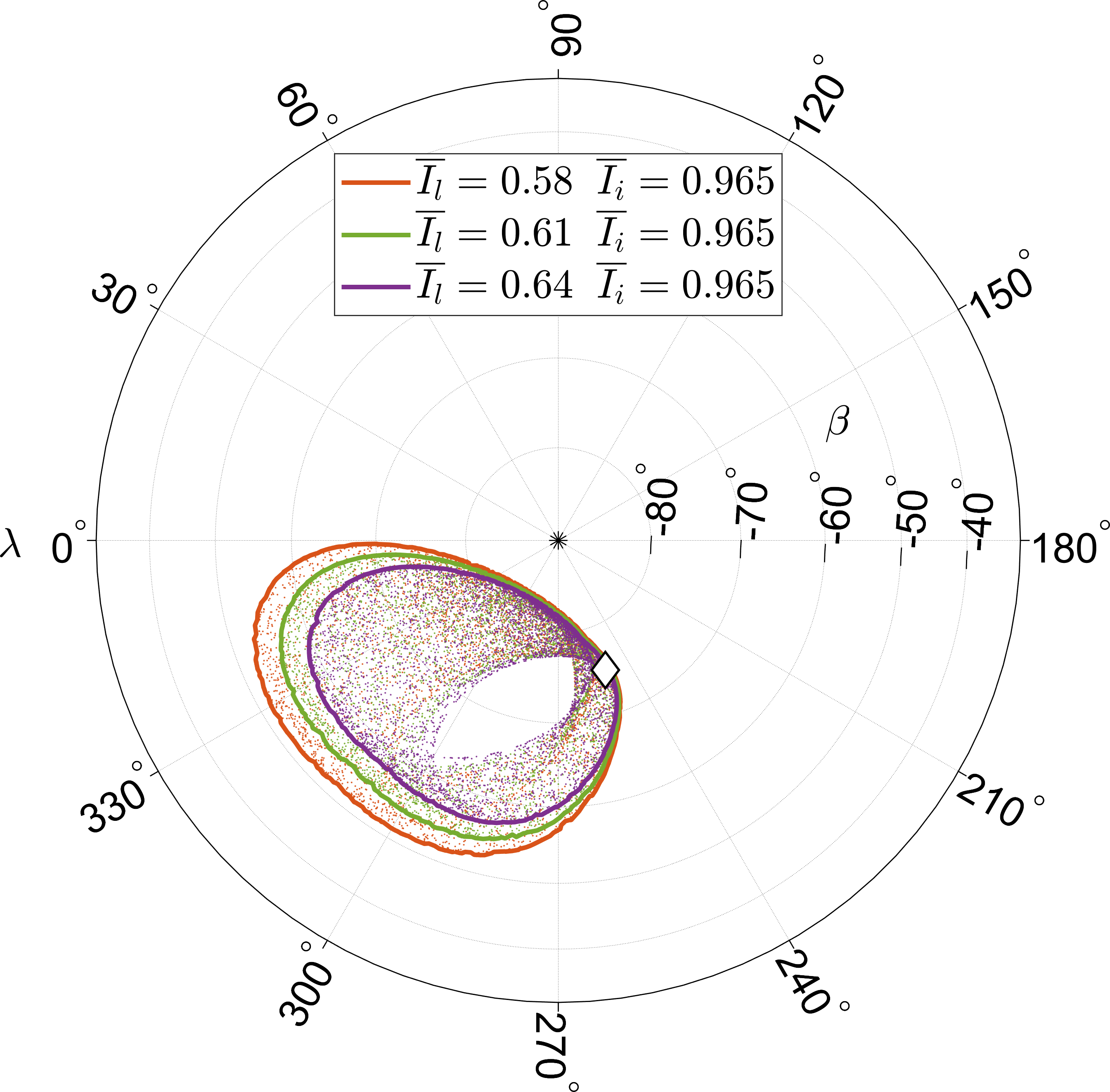}}
    \subcaptionbox{varying $\overline{I_i}=I_i/I_s$}{\includegraphics[width=3.2in]{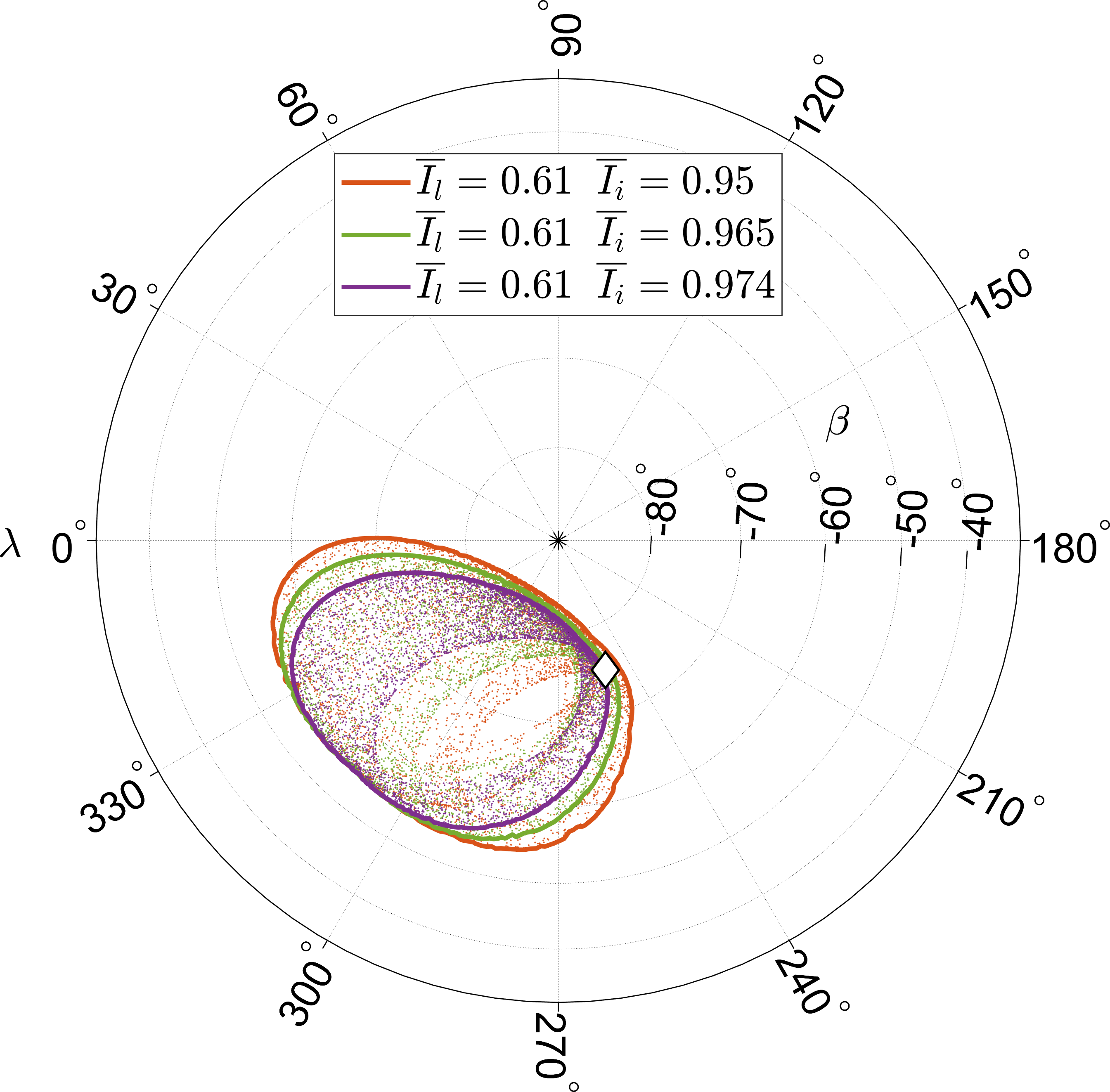}}
    \caption{Post-flyby pole sensitivity to Apophis' moments of inertia assuming the nominal photometric pre-flyby periods and pole (denoted by the white diamond). 5000 runs are conducted for each of the six cases, uniformly sampling the attitude phasing parameters $\phi_o$ and $\tau_o$ over all possible values.}
    \label{fig:polesensitivity}
\end{figure}

To better understand the sensitivity of flyby outcomes to Apophis' mass distribution, Figure \ref{fig:polesensitivity} shows how the post-flyby pole distribution changes with the asteroid's moments of inertia assuming the nominal photometric pole and periods. Here we plot the case for the nominal inertias in green and additional cases in orange and purple. The notable changes in these distributions for different inertias demonstrates the promise of better constraining the inertias by tracking the asteroid's spin state evolution through the flyby. 

\section{Implications}

We will now discuss the impact this abrupt spin state change may have on Apophis' geophysical structure, both on the surface and internally. Using coupled finite element and dynamics models accounting for Apophis' rotation and tidal forces/torques, \cite{hirabayashiapophis} found that surface and internal flyby-induced stress variations would be minimal and unlikely to result in significant geophysical change for a homogeneous body, consistent with the averaged stress analysis given in \cite{holsapple2006tidal}. This is also consistent with earlier studies of the flyby \citep{scheeresapophis} and with detailed analysis using granular mechanics codes \citep{demartini2019using}.

\subsection{Changes in surface conditions across the flyby} 

\begin{figure}[htb]
	\centering\includegraphics[width=5in]{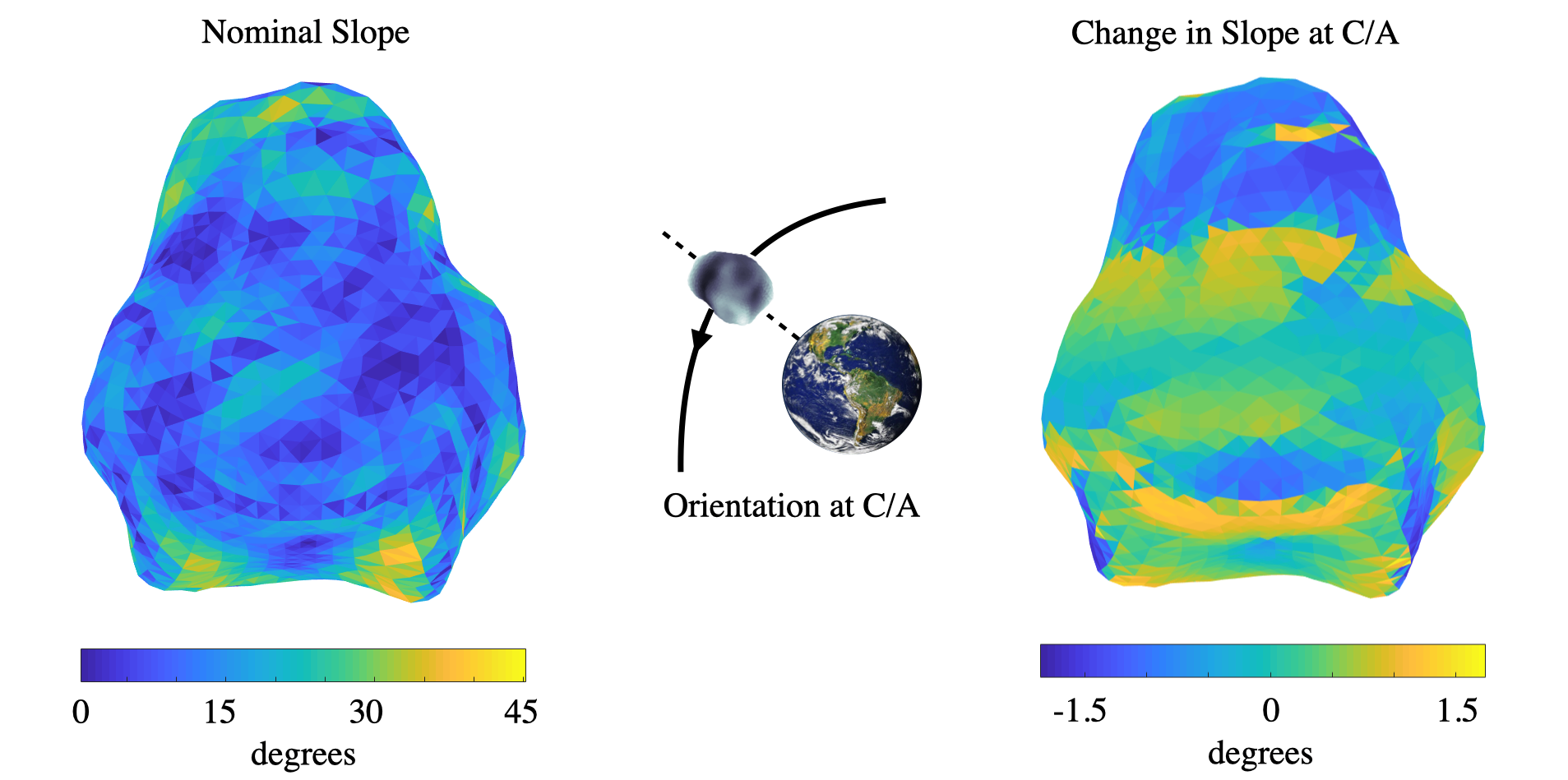}
	\caption{Nominal Apophis slopes (left) and change due to tidal forces at closest approach (right). Here, the bottom of the asteroid faces Earth.}
	\label{fig:slope}
\end{figure}

\begin{figure}[htb]
	\centering\includegraphics[width=5in]{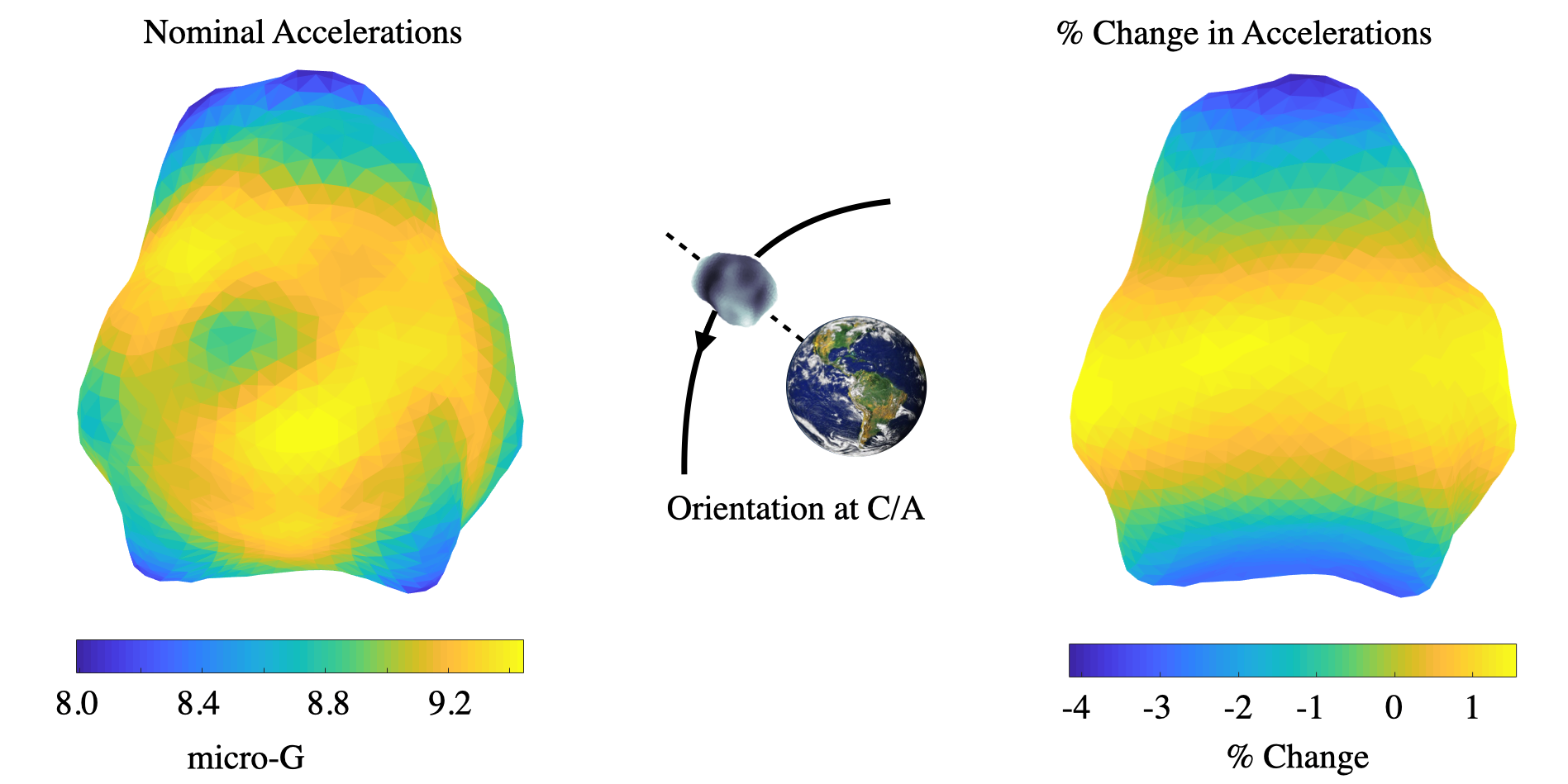}
	\caption{Nominal Apophis surface accelerations (left) and change due to tidal forces at closest approach (right). Here, the bottom of the asteroid faces Earth.}
	\label{fig:accel}
\end{figure}

During the flyby, the asteroid will experience additional forces which will lead to fluctuating surface accelerations that are unlikely in themselves sufficiently large to create observable alterations. The computation of the rotational plus gravity components for the body have been defined in \cite{scheeres_toutatis}. Applying these computations to the constant density radar shape model yield a nominal distribution of surface slopes and surface accelerations. Due to Apophis' complex rotation these values change across the surface periodically. However, given Apophis' low spin rate, these variations are very small compared to self-gravitation. The left plots of Figures \ref{fig:slope} and \ref{fig:accel} show the nominal cases for self-gravitation only, viewed along the maximum inertia axis. Inclusion of spin rate makes minimal changes in these plots. The right plots in Figures \ref{fig:slope} and \ref{fig:accel} show the changes with tidal forces (further discussed later this section). 

During the flyby there are additional perturbations that the asteroid will experience. First are the changes in rotational dynamics due to the torques acting on the body, however these remain small in general as analyzed in detail above. Second is the variation in surface acceleration due to the rotational acceleration of the body, which are simply computed at a location on the body defined by a position vector $\bfm{r}$ as:
\begin{eqnarray}
	\bfm{a}_{acc} & = & - \dot{\bfm{\omega}} \times \bfm{r} 
\end{eqnarray}
where $\dot{\bfm{\omega}}$ is the acceleration of the body's spin rate which, due to the gravitational torques, is much larger than the periodic fluctuation of this term due to its complex spin state. From Figure \ref{fig:wdotca} we see that the maximum magnitude of this acceleration is approximately 4.5 deg/hr$^2$ ($6\times10^{-9}$ rad/s$^2$). Given the average radius of 168 m for the radar shape model, this corresponds to an acceleration on the order of $1\times10^{-6}$ m/s$^2$ (0.1 $\mu$G) or less, which would make a minimal contribution to the nominal surface acceleration of $\sim$8 $\mu$G.

The third and largest contribution will be the direct differential attraction acting on the body due to the Earth's gravity -- commonly termed tidal forces. Assuming that Apophis is small compared to its flyby distance, we can make the usual linear expansion of this difference in attraction to find the tidal acceleration acting on a surface particle relative to the center of the asteroid:
\begin{eqnarray}
	\bfm{a}_{\mathrm{{tide}}} & = & - \frac{\mu_e}{R^3} \left[ \bfm{U} - \hat{\bfm{R}}\hat{\bfm{R}}\right] \cdot \bfm{r} \label{eq:tide}
\end{eqnarray}
Here $\mu_e\sim 398,600$ km$^3$/s$^2$ is the gravitational parameter of the Earth, $R \sim 38,000$ km is the closest approach distance to the Earth, $\bfm{U}$ is the unity dyadic, $\hat{\bfm{R}}\hat{\bfm{R}}$ is the dyadic computed using the unit vectors pointing towards the asteroid from the center of the Earth at closest approach, and $\bfm{r}$ is the position on the surface of the asteroid relative to its center of mass.  The magnitude of the leading term is $\mu_E / R^3 \sim 7.3\times10^{-9}$ s$^{-2}$, which when multiplied by the mean asteroid radius of 168 m (taken from the radar shape model) yields a maximum acceleration of $1.2\times10^{-6}$ m/s$^2$ or 0.12 $\mu$G (less than 2\% of the nominal $\sim$8 $\mu$G surface acceleration). This is comparable to, but larger than, the maximum acceleration from angular acceleration, however that was a clear upper bound. Here, this number is an accurate representation of the differential acceleration that will act across the body. 

In the right plots of Figures \ref{fig:slope} and \ref{fig:accel} we show the change in the surface slope and surface acceleration that will be experienced by Apophis at closest approach, only directly accounting for the tidal acceleration given in Eq.~\ref{eq:tide}. Here we have chosen the orientation of the body that will give the maximum differential accelerations, with the long axis of the asteroid oriented towards the Earth. Given Apophis' relatively slow spin rate, the surface is not near failure conditions. While not considered in this analysis, cohesion of the surface material would further inhibit surface failure. Thus these additional fluctuations around closest approach are not expected to produce significant surface changes assuming a homogeneous body. This was also the conclusion of the earlier analyses in \cite{scheeresapophis}, \cite{demartini2019using}, and \cite{hirabayashiapophis}. We also note the analysis by  \cite{holsapple2006tidal} which estimated that Apophis would need a very low density and a friction angle $\leq$ 5$^{\circ}$ in order for significant resurfacing through closest approach. 

\subsection{Effect of non-homogeneous mass distributions} 

The above analysis treats Apophis as a uniform density body, and focuses on the accelerations occurring at the surface of the body. If we consider a non-homogeneous mass distribution for Apophis then a flyby-induced geophysical change becomes more feasible. A non-homogeneous mass distribution, in this sense, means that the overall asteroid may be separated into larger components that are resting on each other. 

\begin{figure}[H]
    \centering
    \includegraphics[width=4in]{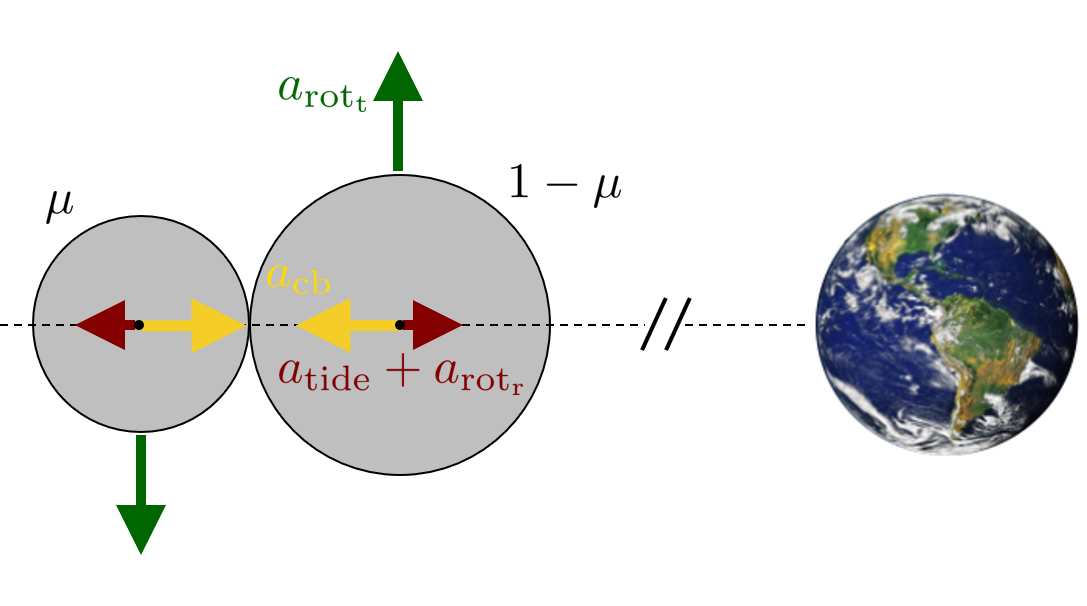}
    \caption{Schematic of Apophis approximated as two spherical components resting on each other with relevant accelerations labeled. We assume the long axis points towards Earth at closest approach. The parameters $\mu$ and $1-\mu$ denote the mass fractions of the two components.}
    \label{fig:contact_binary}
\end{figure}

The delay-Doppler radar images of Apophis suggest the asteroid has a bi-lobed shape. See for example those shown in Figure 2 of \cite{brozovicapophis}. The most extreme model of non-homogeneous mass distribution will model the body as two components resting on each other, maximizing the effect we are studying here and providing a good extreme case for analysis. Previous research has analyzed resting ellipsoids \citep{scheeres_icarus_fission} and contact binary models of general shapes \citep{hirabayashi2019rotationally}. To simplify the current analysis we assume that Apophis consists of two spherical components resting on each other. This simplified model is shown in Figure~\ref{fig:contact_binary} along with relevant accelerations that will be discussed later in this section. These two bodies will have a mutual attraction, creating a compressive force at their point of contact. During closest approach we can evaluate how much this compressive force is reduced due to the tidal forces and rotationally-induced radial acceleration. The combination of relaxing the relative compression combined with the lateral spin acceleration could cause the components to shift, and thus create a change in the mass distribution. 

Assuming Apophis consists of two spherical components resting on each other, the compressive acceleration due to their mutual gravity is simply, 
\begin{equation}
a_{\mathrm{cb}}=\frac{GM}{(r_1+r_2)^2}
\label{eq:acb}
\end{equation}
where $G$ is the gravitational constant,  $M=\frac{4}{3}{\pi}\overline{r}^3\rho$ is the asteroid's total mass, $\overline{r}$ is the asteroid's mean radius taken from the shape model volume, and $\rho$ is the bulk density. The component radii are then $r_1=\mu^{1/3}\overline{r}$ and $r_2=(1-\mu)^{1/3}\overline{r}$ where $\mu$ is the mass fraction of one component. 

Tidal accelerations oppose $a_{\mathrm{cb}}$ (see Figure~\ref{fig:contact_binary}) and will be maximized when the asteroid long axis is pointed towards Earth at closest approach. In this case, the acceleration is, 
\begin{equation}
a_{\mathrm{tide}}=\frac{2\mu_e}{R^3}(r_1+r_2)
\label{eq:atide}
\end{equation}
%

Rotational accelerations due to the tidal torques and spin state, $a_{rot}=\dot{\bm{\omega}}\times\bm{r}+\bm{\omega}\times\bm{\omega}\times\bm{r}$, will also be present. Here $\bm{r}=(r_1+r_2)\hat{\bm{l}}$ is the position vector between the primary and secondary centers of mass. The rotational accelerations can be split into tangential and radial components, $a_{\mathrm{{rot}_t}}=|\dot{\bm{\omega}}\times\bm{r}|$ and $a_{\mathrm{{rot}_r}}=|\bm{\omega}\times\bm{\omega}\times\bm{r}|\sim\omega^2_e(r_1+r_2)$ respectively. These components are shown in Figure~\ref{fig:contact_binary}. 

Assuming $\rho=$ 2 g/cm$^3$, $\overline{r}=$ 168 m, and the well-constrained flyby radius of $\sim$ 38,000 km, we find that the ratio of the tidal acceleration relative to the compressive acceleration is $a_{\mathrm{tide}}/a_{\mathrm{cb}}\leq$ $\sim$ 0.1. Considering also the rotational component with $\omega_e$ corresponding to the pre-flyby $P_{\bar{\phi}}$, $(a_{\mathrm{tide}}+a_{\mathrm{{rot}_r}})/a_{\mathrm{cb}}\leq$ $\sim$0.13. With $a_{\mathrm{cb}}$ proportional to $\rho$, halving the asteroid's density would double these ratios and vice versa. In any case, the components will not separate during the encounter. Then considering the tangential acceleration, $a_{\mathrm{{rot}_t}}/a_{\mathrm{cb}}\leq$ $\sim$0.04 assuming the 4.5 deg/hr$^2$ maximum angular acceleration magnitude from Figure~\ref{fig:wdotca}. Overall, these flyby-induced accelerations first indicate that the two components will remain in contact through the closest approach. However, the reduction of compressive forces between the binary components could create conditions for a shift in the components at closest approach, particularly if there is negligible internal strength to resist ``roll" from the tangential acceleration. Any material in the ``neck" region between the two components would help resist relative motion and provide some cohesive strength. Again, these derived ratios are the maximum possible values which assume: 1) equal mass components (i.e. $\mu=$ 0.5), 2) Apophis' long axis pointed towards the Earth at closest approach, and 3) the largest simulated acceleration.

The scenario for geophysical change implied by this model would consist of two effects, and would be driven by the body's orientation at closest approach. In the following we only consider the previously defined orientation, although future work could generalize the contact binary model and study the possible shifts in more detail. The first effect is a lessening of the contact force between the components, which reduces the sustaining frictional forces that would exist between the resting components. The second is the lateral acceleration due to the angular acceleration, which could then create the conditions for the two components to shift relative to each other, with the simplest motion being a rolling or rocking motion. It is hypothesized that the components of the cometary body 67P have shifted in the past \citep{toshi_nature}. Similar shifts could occur for Apophis, but are certainly not guaranteed and would also require the body to be poised in a favorable way to undergo such a shift. 

\subsection{Detection of mass shifts and internal mass distribution} 

There have been proposals to have a spacecraft in proximity to Apophis around its closest approach to ascertain whether there is any change in the surface or mass distribution of the asteroid. Such a fly-along mission has been analyzed in \cite{scheeres2019stationkeeping}, however it is not necessary to have an in-situ spacecraft perform before and after mappings to detect a change in the mass distribution. If there is a shift in the Apophis components such as described here, the body's moments of inertia will change. Pre and post-flyby spin state solutions provided by ground-based optical and radar observation campaigns will yield moment of inertia ratios. Comparison between the pre and post-flyby values would allow for determination of possible mass shifts. The pre and post-flyby campaigns will also provide independent, high resolution shape model solutions. The shape-derived moments of inertia from the assumed constant density shape models can also be compared to the dynamical moments of inertia obtained from spin state estimation. Differences between the shape-derived and dynamical inertias could provide evidence for density inhomogeneity. The eventual visit of the OSIRIS-APEX spacecraft to Apophis \citep{orex_apophis} will further refine the asteroid's post-flyby spin state and mass distribution for comparison with the pre-flyby values. OSIRIS-APEX's detailed surface mapping operations will also allow for determining whether there are freshly uncovered surfaces due to regolith displacement.

Finally, the measured change in spin state across the flyby, independent of possible changes in the moments of inertia, can also be used to model the asteroid's mass distribution more precisely. Here, the spin dynamics and gravitational torques are combined and provide unique insight on the second degree and order mass moments of Apophis. A similar analysis as performed by \cite{takahashi2013spin} for Toutatis can be conducted, with the mass distribution sensitivity for Apophis being much greater given its closer encounter distance.

\subsection{Yarkovsky Acceleration}

\begin{figure}[H]
    \centering
    \includegraphics[width=4in]{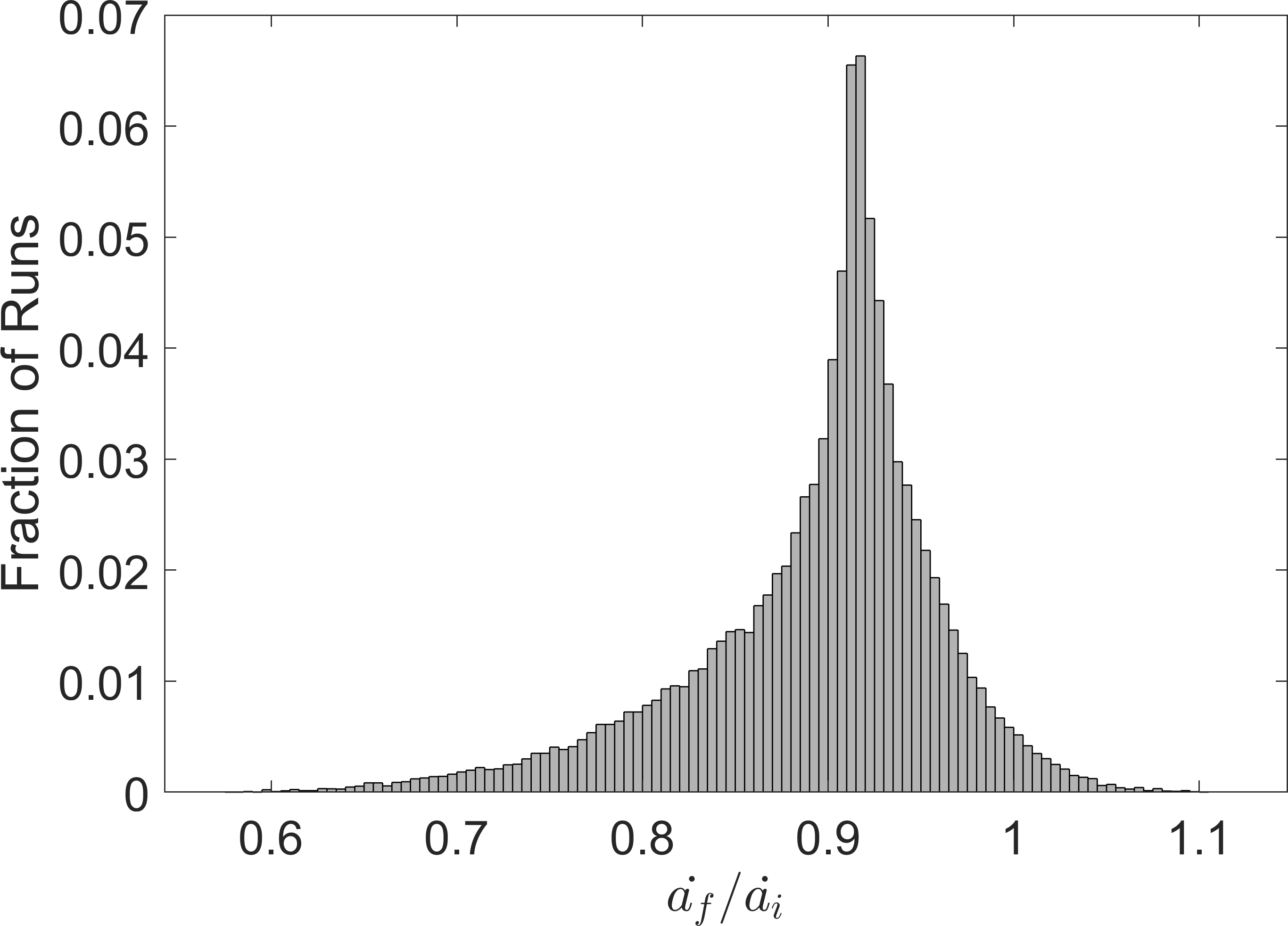}
    \caption{Distribution of the relative post-flyby Yarkovsky semi-major axis drift rate.}
    \label{fig:adotratio}
\end{figure}

Flyby-induced changes to Apophis' heliocentric orbit and spin state will also affect the asteroid's Yarkovsky acceleration and resulting semi-major axis drift rate. For small asteroids like Apophis, the diurnal drift component tends to be significantly larger than the seasonal component \citep{asteroidsIVyarkovskyyorp}. So here we only consider the diurnal component. We assume that the changing spin periods have a negligible effect on the drift rate. Furthermore, studies have shown that the Yarkovsky drift models for uniform rotators yield accurate approximations for tumblers as well \citep{asteroidsIVyarkovskyyorp}. So using the diurnal equation for uniform rotators and canceling out all constant terms, the ratio of the post-flyby semi-major axis drift rate $\dot{a_f}$ to the pre-flyby value $\dot{a_i}$ can be written as \citep{asteroidsIVyarkovskyyorp},
\begin{equation}
\frac{\dot{a_f}}{\dot{a_i}}=\frac{\cos{\gamma_f}}{\cos{\gamma_i}}\sqrt{\frac{a_i}{a_f}}
\label{eq:adotratio}
\end{equation}
where $a_i$ and $a_f$ are the corresponding semi-major axes and $\gamma$ is the angle between the rotational and heliocentric orbit angular momentum vectors. The pre and post-flyby orbit poles were calculated from JPL Horizons ephemerides. Accounting for the different pre and post-flyby rotation pole directions for each run, the resulting distribution of the relative post-flyby drift rate is provided in Figure~\ref{fig:adotratio}. The semi-major axis will increase from 0.92 AU to 1.10 AU, so $\sqrt{a_i/a_f}=$ 0.91. The shape of the distribution is driven by $\cos{\gamma_f}/\cos{\gamma_i}$. For 56\% of runs, $\cos{\gamma_f}/\cos{\gamma_i}$ less than 1, indicating the majority of runs have post-flyby poles further from the orbit south pole. Values for $\cos{\gamma_f}/\cos{\gamma_i}$ range from 0.63 to 1.2, so the drift rate does not change sign. In all, Figure~\ref{fig:adotratio} shows that the minimum and maximum post-flyby drift rates are 57\% and 112\% of the pre-flyby value respectively, with 97\% of runs yielding a reduction in drift rate. Again, this is due to both the flyby-induced increase in semi-major axis and the general shift in post-flyby rotation poles away from the orbit south pole.

\section{Conclusions}

Apophis’ tumbling periods, excitation level, and inertial pole direction will be significantly altered due to the 2029 encounter. The resulting spin state evolution is very sensitive to the asteroid’s closest approach attitude and mass distribution. Sensitivity to the latter will aid in refining the asteroid’s center of mass and moments of inertia through pre and post flyby tracking. Peak accelerations obtained from the flyby simulations indicate that resurfacing and internal distortion are unlikely. Nevertheless, if Apophis is a contact binary, the components could potentially shift at closest approach. This deformation may be detectable through remote and in-situ observation. For example, tracking the complex rotation state prior to and after the flyby could be used to detect changes in the body's mass distribution. In the contact binary case, any observed shifts would inform our understanding of contact binary structure and evolution.  Flyby-induced changes to Apophis' semi-major axis and rotation pole are also likely to decrease the asteroid's Yarkovsky acceleration by tens of percent. Precise orbit determination enabled by the OSIRIS-APEX rendezvous will help estimate the post-flyby Yarkovsky drift rate. While a small perturbation, this semi-major axis drift rate is relevant for long-term orbit prediction and mapping out post-2029 Earth encounters.

The biggest implication of this work is that any mission planning should account for Apophis' post-encounter spin state which could be very different from its current state. Prediction of this post-flyby state is very important given the short timespan between the April 2029 encounter and OSIRIS-APEX's arrival later that year. This is further stressed by the fact that Apophis will remain in non-principal axis rotation after the flyby. Such states generally require longer observation arcs to estimate compared to uniform rotation, especially given Apophis' multi-day long axis rotation. Surface mapping, relative navigation, and regolith excavation efforts will depend heavily on accurate spin state estimates. Overall, this illustrates the importance of obtaining accurate pre-encounter spin state and mass distribution estimates from ground-based and possibly in-situ spacecraft observations. 

\section*{Acknowledgements}
\noindent

The authors would like to thank Paul Abell and Richard Binzel for their insightful questions and suggestions which significantly enhanced this paper. The work by CJB and DJS was supported by NASA grant 80NSSC22K0240. The work done by M. Brozovi\'{c} and S. Chesley was carried out at the Jet Propulsion Laboratory, California Institute of Technology, under contract with the National Aeronautics and Space Administration (80NM0018D004). The work by P. Pravec and P. Scheirich was supported by the Grant Agency of the Czech Republic, Grant 20-04431S.  

\bibliography{references}

\end{document}